\documentclass{article}
\usepackage{mpla}
\usepackage{psfig}
\begin{document}
\setlength{\textheight}{7.7truein}  

\runninghead{Sakharov's induced gravity: a modern perspective.}
{Sakharov's induced gravity: a modern perspective}

\normalsize\textlineskip
\thispagestyle{empty}
\setcounter{page}{1}

\copyrightheading{}			

\vspace*{0.88truein}

\fpage{1}
\centerline{\bf SAKHAROV'S INDUCED GRAVITY:}
\baselineskip=13pt
\centerline{\bf A MODERN PERSPECTIVE}
\vspace*{0.37truein}
\centerline{\footnotesize MATT VISSER\footnote{
Permanent address after 1 July 2002:
\\ 
School of Mathematics and Computer Science, Victoria University, PO Box 600, 
Wellington, New Zealand.}}
\baselineskip=12pt
\centerline{\footnotesize\it Physics Department, Washington University in Saint Louis}
\baselineskip=10pt
\centerline{\footnotesize\it Saint Louis, Missouri 63130-4899, USA}

\vspace*{0.225truein}

\publisher{(15 April 2002)}{}

\vspace*{0.21truein}
\abstracts{
Sakharov's 1967 notion of ``induced gravity'' is currently enjoying a
significant resurgence.  The basic idea, originally presented in a
very brief 3-page paper with a total of 4 formulas, is that gravity is
not ``fundamental'' in the sense of particle physics.  Instead it was
argued that gravity (general relativity) emerges from quantum field
theory in roughly the same sense that hydrodynamics or continuum
elasticity theory emerges from molecular physics. In this article I
will translate the key ideas into modern language, and explain the
various versions of Sakharov's idea currently on the market. }{}{}

\bigskip
\textlineskip
\noindent
Contribution to the ``{\emph{First IUCAA Meeting on the Interface of
Gravitational and Quantum Realms}'', held in Pune in December 2001.
\\
To appear in Modern Physics Letters A; \hfill arXiv: gr-qc/0204062.



\def\Str{\hbox{Str}}
\def\Tr{\hbox{Tr}}
\def\str{\hbox{str}}
\def\tr{\hbox{tr}}
\def\implies{\Rightarrow} 
\def\d{{\mathrm{d}}}
\def\be{\begin{equation}}
\def\ee{\end{equation}}
\def\bea{\begin{eqnarray}} 
\def\eea{\end{eqnarray}}  
\def\alt{\mathrel{\mathpalette\vereq{<}}}%
\def\lesssim{\mathrel{\mathpalette\vereq{<}}}%
\def\agt{\mathrel{\mathpalette\vereq{>}}}%
\def\gtrsim{\mathrel{\mathpalette\vereq{>}}}%
\def\vereq#1#2{%
 \lower3pt\vbox{%
  \baselineskip1.5pt 
  \lineskip1.5pt
  \ialign{$#1\hfill##\hfil$\crcr#2\crcr\sim\crcr}%
 }%
}%
\setcounter{footnote}{0}
\renewcommand{\thefootnote}{\alph{footnote}}
\vspace*{1pt}\textlineskip	
\section{Introduction}	
\vspace*{-0.5pt}
\noindent
Einstein gravity (general relativity) is based on two key elements:
(1) pseudo-Riemannian geometry ({Lorentzian geometry}) --- the
geometric notions derived from the Einstein Equivalence Principle, and
(2) specific field equations for the Ricci tensor.  It is the second
of these elements that Sakharov addresses in developing his notion of
``induced gravity''.\cite{Sakharov} There are by now a number of
versions of ``induced gravity'' on the market, some of which focus
primarily on particle physics and others which focus primarily on
(semiclassical) gravity.

To set the general framework:
\begin{itemize}
\item
Assume you are given a Lorentzian manifold.
\item
Make no assumptions about the dynamics of this geometry; leave it free
to flap in the breeze. Do not attempt to quantize geometry/gravity
itself, but quantize everything else. (The geometry is considered as a
classical background.)
\item
Consider one-loop quantum field theory on this manifold.
\item
Then at one loop the effective action is guaranteed to contain terms
of the form:
\be
\int \d^4x \sqrt{-g} 
\left\{ 
c_0 + c_1\; R(g) + c_2 \; (\hbox{``}R^2\hbox{''}) 
\right\}.
\ee
Compare this with the standard Lagrangian for Einstein gravity
\be
\int \d^4x \sqrt{-g} 
\left\{ 
-\Lambda - {R(g)\over16\pi\; G}  + K \; (\hbox{``}R^2\hbox{''}) 
+ {\cal L}_{\mathrm{matter}}
\right\}.
\ee
That is: the one loop effective action automatically contains terms
proportional to the cosmological constant, the Einstein--Hilbert
action, plus ``curvature-squared'' terms.
\end{itemize}
This is the central observation, and it is extremely suggestive, but
is it enough for us to recover all the interesting aspects of gravity?
It is the answer to this last question that generates all the
interest, and all the difficulties and subtleties.

\section{One-loop matter affects classical gravity}
\noindent
Start by considering the one-loop contribution to the effective action
for a (possibly non-minimally coupled) scalar field:
\be
{\cal S}_g 
= -{1\over2}\ln\det(\Delta_g + m^2 + \xi R) 
=  -{1\over2}\Tr\ln(\Delta_g + m^2 + \xi R).
\ee
In Feynman diagram language this determinant represents the sum of all
one-loop diagrams coupled to an arbitrary number of classical external
gravitons. (Even though we have not at this stage defined gravity, we
have defined the notion of geometry. To connect with particle physics
language we temporarily assume that the geometry is close to flat,
expand the spacetime metric as $g_{ab} = \eta_{ab} + h_{ab}$, and
Fourier transform the $h_{ab}(x)$. The resulting $h_{ab}(k)$ are
defined to be ``external gravitons''. Once you have developed this
Feynman diagram picture, purely as an aid to visualization, there is
no need to remain close to flat space as the determinant formula
continues to make excellent sense even for highly curved and/or
topologically nontrivial manifolds.) For Feynman diagrams whose only
external legs are classical gravitons, particle-particle interactions
will contribute only at two-loops or higher.

The determinant can be rigorously defined by a number of different
techniques (e.g, zeta functions~\cite{zeta,BVW}). It is somewhat more
useful to adopt an explicit cutoff.  Use the standard identity
\be
\ln(b/a) = \int_0^\infty {\d x\over x} \left[ e^{-ax} - e^{-bx} \right].
\ee
Now define $\Tr$ to include a trace over both spacetime and any
internal indices
\be
\Tr[\ \ ] \equiv \int \d^4 x \; \tr [\ \ ]
\ee
and adopt Schwinger's proper time formalism.\footnote{%
There is nothing particularly special about Schwinger's regularization
--- for instance, we could just as easily use Pauli--Villars
regularization. On the other hand, zeta functions or dimensional
regularization are a trifle tricky because they are in a sense too
powerful, and have the effect of ``hiding'' some of the interesting
terms. }
\ This yields the simple expression
\be
{\cal S}_g  = {\cal S}_{g_0} + 
{1\over2}\Tr \int_0^\infty {\d s\over s} 
\left[ \exp(-s [\Delta_g + m^2 + \xi R] )  
- \exp(-s [\Delta_{g_0} + m^2 + \xi R_0] ) \right].
\ee
Note that this computes the \emph{difference} in the one-loop
contribution to the effective that comes from comparing two different
metrics ($g$ and $g_0$) defined on the \emph{same} topological
manifold. ($g_0$ is simply any convenient reference metric.)  Since
this expression is ultraviolet divergent, one should regularize with a
short distance cutoff. Note that dimensionally $[s] = M^{-2} =
L^2\equiv [\kappa^{-2}]$. The regularized but unrenormalized one-loop
contribution to the effective action is then
\be
{\cal S}_g  = {\cal S}_{g_0} + 
{1\over2}\Tr \int_{\kappa^{-2}}^\infty {\d s\over s} 
\left[ \exp(-s [\Delta_g + m^2 +\xi R] ) 
- \exp(-s [\Delta_{g_0} + m^2+ \xi R_0] ) \right].
\ee
Now use the heat kernel expansion as $s\to 0$:\cite{zeta,BVW}
\be
\exp(-s [\Delta_g + m^2 + \xi R] ) =  
{\sqrt{-g}\over(4\pi\,s)^2} \left[ a_0(g) + a_1(g) s + a_2(g) s^2 + \cdots \right].
\ee
The coefficients $a_n$ are variously denoted the Seeley--DeWitt
coefficients, Hami\-dew coefficients, or the Minakshisundarum--Pliejel
coefficients.~\cite{BVW} They are universal functions of the spacetime
geometry, and we do not need to know their explicit form just yet.
Then
\begin{eqnarray}
{\cal S}_g  &=& {\cal S}_{g_0} + 
{1\over32\pi^2}\Tr\Bigg\{ [a_0(g) - a_0(g_0)] \;{\kappa^4\over2}  
+[a_1(g) - a_1(g_0)] \; {\kappa^2}
\nonumber 
\\
&&+  [a_2(g) - a_2(g_0)] \; {\ln (\kappa^2/m^2)} \Bigg\} 
+\hbox{UV finite.} 
\end{eqnarray}
That is, the use of the heat kernel expansion has permitted us to
isolate the both the form and severity of potential divergences --- a
result that is of course well known and is the basis, in one form or
another, of essentially all work in curved-space quantum field
theory.~\cite{B&D} (For that matter, the occurrence of quartic,
quadratic, and logarithmic divergences is quite standard in flat-space
QFT.)

If we now consider Dirac/Weyl spinors, then provided there is no net
chiral anomaly, we can square the Dirac operator to write
\bea
{\cal S} &=& + \ln\det([\gamma\cdot D]+m) 
\nonumber\\
&=& +{1\over2}\ln\det(-[\gamma\cdot D]^2+m^2)
\nonumber\\
&=& +{1\over2}\ln\det\left(\Delta + {{1\over4} R} +m^2\right).
\eea
Note the relative minus sign; in fact all fermions contribute with a
relative minus sign.  Summing over all particles, bose plus fermi:
\begin{eqnarray}
{\cal S}_g  &=& {\cal S}_{g_0} + 
{1\over32\pi^2}\Str\Bigg\{ [a_0(g) - a_0(g_0)] \; {\kappa^4\over2}
+[a_1(g) - a_1(g_0)] \; {\kappa^2} 
\nonumber\\
&&+  [a_2(g) - a_2(g_0)] \; {\ln (\kappa^2/m^2)} \Bigg\} 
+\hbox{UV finite.} 
\end{eqnarray}
Here we have introduced the ``supertrace'' operator $\Str$, which sums
over all particle species but weights fermi fields with a relative
minus sign. Specifically
\be
\Str[\ \ ] \equiv \Tr\left[ (-)^F \ \ \right].
\ee
Note that the usefulness of this ``\Str'' object is {\emph{not}}
limited to supersymmetric theories, it makes perfectly good sense in
arbitrary QFTs, and is simply a useful notational and bookkeeping
trick.

Working from any of many standard references it is easy to see\cite{B&D}
\be
a_0(g) = 1.
\ee
\be
a_1(g) =   k_1 \;R(g) - m^2.
\ee
\bea
a_2(g) &=& 
k_2 \;C_{abcd} C^{abcd} + k_3 \;R_{ab} R^{ab} + k_4 \;R^2 + k_5 \;\nabla^2 R 
\nonumber
\\
&&\qquad\qquad\qquad - m^2 \; k_1 \;R(g) + {1\over2} m^4. 
\eea
where the dimensionless constants $k_i$ depend on the particular
particle species being considered. The form of these coefficients can
be deduced by power counting and the fact that they must be scalar
invariants of the spacetime geometry. The particular coefficients
involving the $m^2$ and $m^4$ terms are easily deduced by combining a
Taylor expansion of $\exp(-s\; m^2)$ with the corresponding $m=0$
Seeley--DeWitt expansion.  Not all of the terms in $a_2$ are truly
independent.  Upon integration
\be
\int \sqrt{-g} \; a_2(g) = 
\int \sqrt{-g} \; \left\{ k_2' \;C_{abcd} \; C^{abcd} + k_4' \; R^2
- m^2 k_1 \; R(g) + {1\over2} m^4 \right\}.
\ee
The $k_5$ term has been discarded because it is a total divergence,
while the $k_3$ term can be eliminated using the integral formula for
the 4-dimensional Euler characteristic. $k_2'$ and $k_4'$ are suitable
linear combinations of $k_2$, $k_3$, and $k_4$. For most of the
qualitative discussion below we will not need to know specific values
for the $k_i$.

Suppose we now define $\str$ by
\be
\Str [\ \ ] \equiv \int d^4 x \;\; \str [\ \ ]
\ee
and proceed to unwrap the above results by collecting terms based on
their geometrical form:
\begin{eqnarray}
{\cal S}_g  &=& {\cal S}_{g_0} + 
{1\over32\pi^2}\Bigg\{ 
{\str\left[{\kappa^4\over2} - m^2 \kappa^2 
+ {m^4\over2} \ln\left({\kappa^2\over m^2}\right) \right]}
\int \d^4x \left[\sqrt{-g} - \sqrt{-g_0} \right]
\nonumber\\
&&+  {\str\left[k_1 \kappa^2 - k_1 m^2 \ln\left({\kappa^2\over m^2}\right) \right] }
\int \d^4x \left[\sqrt{-g} \; R(g) - \sqrt{-g_0} \; R(g_0) \right]
\nonumber\\
&&+  {\str\left[k_2' \ln \left({\kappa^2\over m^2}\right)\right]}
\int \d^4x \left[\sqrt{-g} \; C^2(g) - \sqrt{-g_0} \; C^2(g_0) \right]
\nonumber\\
&&+  {\str\left[k_4' \ln \left({\kappa^2\over m^2}\right)\right]} 
\int \d^4x \left[\sqrt{-g} \; R^2(g) - \sqrt{-g_0} \; R^2(g_0) \right]
\Bigg\} 
\nonumber\\
&&+ \hbox{UV finite.} 
\end{eqnarray}
It is now trivial to extract the coefficients in the one-loop effective
action. Define the gravitational couplings as follows
\be
\int \d^4x \sqrt{-g} 
\left\{ 
-\Lambda - {R(g)\over16\pi\; G}  + K_2 \; C_{abcd} \; C^{abcd} + K_4 \; R^2 
+ {\cal L}_{\mathrm{matter}}
\right\}.
\ee
Then explicitly keeping track of potential zero-loop (tree level)
contributions\footnote{%
Of course, admitting the possibility of zero-loop contributions is
anti-Sakharov in spirit, but we shall see uses for this abberant
behaviour later.}
%
\be
\Lambda = 
\Lambda_0 
- {1\over32\pi^2}\;
\str\left[
{\kappa^4\over2} - m^2 \kappa^2 + {m^4\over2} \ln\left({\kappa^2\over m^2}\right) 
\right]
+ \hbox{UV finite.} 
\ee
\be
{1\over G} = {1\over G_0} 
- {1\over2\pi}\;
\str\left[k_1 \kappa^2 - k_1 m^2 \ln\left({\kappa^2\over m^2}\right) \right] 
+ \hbox{UV finite.} 
\ee
\be
K_2 = (K_2)_0 + {1\over32\pi^2}\;
\str\left[k_2' \ln \left({\kappa^2\over m^2}\right)\right]  
+ \hbox{UV finite.} 
\ee
\be
K_4 = (K_4)_0 + {1\over32\pi^2}\;
\str\left[k_4' \ln \left({\kappa^2\over m^2}\right)\right] 
+ \hbox{UV finite.} 
\ee
These are \emph{regulated} but \emph{unrenormalized} one-loop
expressions describing the influence of the matter sector on the
gravitational couplings.  We see that armed with a knowledge of the
particle spectrum and the cutoff, the one-loop QFT-induced changes in
the gravitational couplings are calculable.  With these formulae in
hand, it is now easy to go back to Sakharov's paper\cite{Sakharov}
and see where his ideas were coming from. It is also important to note
that it is at this stage that the road diverges. There are at least
four main directions one can take ---
\begin{itemize}
\item
Sakharov: Demand one-loop dominance.
\item
Pauli: Demand one-loop finiteness.
\item
Frolov--Fursaev: Demand one-loop calculability.
\item
Renormalizability: Demand one-loop calculability for certain
\emph{differences}.
\end{itemize}
The same one-loop formulae can then have rather different implications
depending on the route you take. After a brief digression on
observational bounds, let us consider the major options in turn.

\section{Experimental/Observational bounds}

Regarding the gravitational couplings, we have good experimental data
on Newton's constant $G$, while for the cosmological constant we are
certainly safe in asserting
\be
|8\pi G \; \Lambda | \alt  10^{-120} \; M_{\mathrm{Planck}}^4.
\ee
Current observational data indeed favours the stronger statement
\be
8\pi G \; \Lambda  \approx + 10^{-123} \; M_{\mathrm{Planck}}^4.
\ee
In contrast, constraints on the dimensionless numbers $K_2$ and $K_4$
are quite weak (amazingly weak). The linearized theory leads to a
weak-field potential of Yukawa form\cite{Stelle}
\be
\phi(r) \propto {1\over r} 
- {4\over3} {\exp(-M_{\mathrm{Planck}}\,r/\sqrt{K_2})\over r}
+ {1\over3} {\exp(-M_{\mathrm{Planck}}\,r/\sqrt{K_4})\over r}.
\ee
Note that the strength of the Yukawa terms is fixed and it is only
their ranges that depend on $K_{2/4}$.

Negative values for $K_2$ and $K_4$ lead to objectionable tachyonic
behaviour. (For instance, the static potential would not limit to the
Newtonian value at large distances but would instead oscillate wildly
between zero and $8/3$ times the standard Newtonian value. In addition
the linearized theory then exhibits various
instabilities.\cite{Horowitz})

Positive values for $K_2$ and $K_4$ are constrained by the fact that
we have directly tested the inverse square law for gravity down to
approximately the millimeter scale, implying
$M_{\mathrm{Planck}}/\sqrt{K} \gtrsim 10^{-13} \hbox{ GeV}$. That is
\be
K_{2/4} \lesssim 10^{+64}.
\ee
This direct experimental bound leaves considerable maneuvering
room. (Various ``fifth force'' experiments currently in progress are
likely to improve these bounds somewhat in the not too distant
future.\cite{Adelberger}) Indirect arguments, based on using the $K_4$
term to drive cosmological inflation suggest\cite{Suen}
\be
K_4 \in (10^{+11}, \, 10^{+15} ).
\ee
(Cosmology is rather insensitive to the $K_2$ term since the Weyl
tensor is identically zero in all FRW backgrounds.)  There is
certainly considerable room for deriving better phenomenological and
observational bounds on these parameters.

\section{Sakharov: one-loop dominance}
\noindent
Sakharov's own interpretation of the manner in which one-loop matter
influences gravity was this:\cite{Sakharov}
\begin{itemize}
\item
Set all tree-level constants to zero. 
\item
Assume one-loop physics is dominant.
\item
Assume most dimensionless numbers of order one.
\item
Assume ${\kappa \approx M_\mathrm{Planck}}$; so there is an explicit
cutoff at the Planck scale.
\item
Quietly agree to ignore $\Lambda$, $K_2$, and $K_4$.
\end{itemize}
Then the Newton constant is ``induced'' at one-loop. Keeping only the
dominant terms in the divergence
\be 
{1\over G} \approx 
-{1\over2\pi}\;\str[k_1] \; \kappa^2; \qquad \str[k_1] \approx -1.
\ee
This is the key observation: an approximate formula for Newton's
constant as a function of the cutoff $\kappa$ and selected features of
the particle spectrum (encoded in $\str[k_1]$).  It is this
conjectured connection between Newton's constant and particle physics
that has led to so much attention being given to this idea. (Over 260
citations as of February 2002.)  This route certainly paints a
coherent and attractive physical picture --- but is there real
predictive power? And is it possible to say more?

For instance, if we try to say something about the other gravitational couplings:
\be
\Lambda \approx  
-{1\over64\pi^2}\;\str[I] \; \kappa^4; \qquad \str[I] \approx 0.
\ee
\be
K_2 \approx  {1\over32\pi^2} \;\str[k_2'] \;\ln (\kappa^2/\mu^2) \approx 1. 
\ee
\be
K_4 \approx   {1\over32\pi^2} \; \str[k_4'] \;\ln (\kappa^2/\mu^2)  \approx 1. 
\ee
Note that powers dominate over logarithms (wherever possible), and
that the smallness of the observed cosmological constant is very much
put in by hand. (More on this below.) In contrast $K_2$ and $K_4$ are
naturally dimensionless numbers of order unity; which is certainly
compatible with experiment. 

\section{Pauli: QFT compensation}
\noindent
A different interpretation of (some of) these formulae can be traced
back all the way to Wolfgang Pauli.  The key discussion is contained
in his 1950 lectures on quantum field theory (the
``Feldquantisierung'').\cite{Pauli} Because Pauli was working in flat
space, not curved, his comments were directed towards the cosmological
constant $\Lambda$ (which he rephrased in terms of the net zero-point
energy of the QFT).\footnote{%
One particularly transparent way of seeing that the shift in the
cosmological constant as formulated above is equivalent to the net
zero point energy is to note that\cite{Pauli}
\[
\int_0^\kappa k^2 \sqrt{m^2+k^2} dk = 
{\kappa^4\over4} + {m^2\kappa^2\over4} - {m^4\; \ln(\kappa^2/m^2)\over8}
+ \hbox{finite}.
\]
The left hand side is manifestly proportional to the zero point
energy, while the right hand side precisely yields the one-loop shift
in $\Lambda$.}
\ Pauli did not himself consider Newton's constant $G$, but the relevant
extensions are straightforward.

\underline{Step 1:} 
To guarantee a one-loop finite result for $\Lambda$ you
need:\cite{Pauli}
\be
\str(I) = \str(m^2) = \str(m^4) = 0.
\ee
Once these finiteness constraints are enforced, the one-loop
contribution to the cosmological constant is\footnote{%
Indeed, as Pauli points out, if you now add the additional restriction
$\str[m^4\;\ln(m^2/\mu^2)]$ then the one loop contribution to the
cosmological constant (zero point energy) is not just finite, it's
zero.}
\be
{\Lambda = 
\Lambda_0 
+ {1\over64\pi^2}\;\str\left[m^4 \ln\left({m^2\over\mu^2}\right) \right]
+ \hbox{two loops.}} 
\ee
These finiteness constraints are a fore-runner of the idea of
supersymmetry, and were certainly known to many of the developers of
supersymmetry (SUSY).  They give tight constraints on the particle
physics content of the model; constraints that are certainly
{\emph{not\/}} satisfied by the standard model (SM), and certainly
require ``beyond the standard model'' (BSM) physics.  For instance,
these constraints are satisfied by arbitrary SUSY theories prior to
spontaneous symmetry breakdown (SSB), and by all of the non-SUSY
one-loop finite QFTs.\cite{finite} Note the quantity $\mu$ appearing
here is simply any convenient mass scale chosen to make the argument
of the logarithm dimensionless, and that $\Lambda$ is independent of
$\mu$ thanks to the constraint $\str(m^4)=0$.

\underline{Step 2:} 
Now extend Pauli's compensation idea to curved space. If you
additionally assume
\be 
\str(k_1) = \str(k_1 \; m^2) = 0,
\ee
then the one-loop contribution to Newton's constant is finite and
\be
{{1\over G} = 
{1\over G_0} 
- {1\over2\pi}\;\str\left[k_1 \;m^2 \;\ln\left({m^2\over\mu^2}\right) \right]
+ \hbox{two loops.}}
\ee
Note that the finiteness constraint $\str(k_1)=0$ is completely at
odds with Sakharov's original version of the induced gravity
proposal. Also note that this constraint guarantees that $G$ is
independent of $\mu$.

Arranging these secondary finiteness constraints is considerably more
complicated, the relevant coefficients being given in Table I.  The
vanishing of $\str(k_1)$ and $\str(k_1\;m^2)$ are now \emph{very}
strong constraints on the particle content. These constraints are not
derivable from unbroken SUSY alone, nor does one necessarily need SUSY
to satisfy these constraints. A suitable toy model has been presented
by Frolov and Fursaev.\cite{Frolov}

\underline{Step 3:} 
For the final stage, you need to additionally assume
\be
\str(k_2') = \str(k_4') = 0
\ee
in order to render the one-loop contributions to the ``$R^2$''
coupling constants finite. Under this assumption
\be
K_{2/4} = 
(K_{2/4})_0 
- {1\over32\pi^2} \;
\str\left[ k_{2/4}' \; \ln\left({m^2\over\mu^2}\right) \right]
+ \hbox{two loops.} 
\ee
Note that $k_2'$ and $k_4'$ are rather messy coefficients that seem to
follow no particular pattern. The quantities in Table I were generated
from the corresponding results presented in Birrell and
Davis.\cite{B&D} (No explicit toy model satisfying these Step 3
constraints in addition to Steps 1 and 2 is currently known; it
appears that one would need to include higher-spin fields.)

\begin{table}[htbp]
\tcaption{The coefficients $k_0$, $k_1$, $k_2'$ and $k_4'$.}
\centerline{\footnotesize\smalllineskip
\begin{tabular}{l c c c c}\\
\hline
{Description} & {$k_0$} & {$k_1$} & $k_2'$ & $k_4'$\\
\hline
Scalar (minimal)   & $1$ & ${1/6}$ & ${1/120}$ & $1/72$ \\
Scalar (conformal) & $1$ & $0$ & ${1/120}$ & $0$ \\
Scalar (generic)   & $1$ & ${1/6}-\xi$ & ${1/120}$ & $(1-6\xi)^2/72$ \\
Weyl spinor        & $2$ & $-{1/6}$ & $-{1/40}$ & $0$ \\
Dirac spinor       & $4$ & $-{1/3}$ & $-{1/20}$ & $0$ \\
Vector($1\oplus0$) & $4$ & ${-1/3} $ & ${7/60}$ & ${1/36}$\\
Spin 1 (massive)   & $3$ & $-{1/2}$&  ${13/120}$ & 1/72 \\
Spin 1 (massless)  & $2$ & $-2/3$ & 1/10 & 0 \\
\hline
Chiral supermultiplet            & $0$ & 1/2 & 1/24 & 1/36 \\
Vector supermultiplet (massless) & $0$ & 1/2 & 1/8 & 0\\
Vector supermultiplet (massive)  & $0$ & 0 & 5/24 & 1/24 \\
\hline\\
\end{tabular}}
{\footnotesize
Total contributions to the $k_i$ summed over spin states for low-spin
particles. (Here $k_0 = \tr[I\,]$ counts the number of spin states.) For
supermultiplets the \emph{net} value, including the effect of the
minus sign for fermions, is reported.}
\end{table}

The overall conclusion is that one can keep all the one-loop effects
of matter on the gravity sector finite, at the cost of very strong
constraints on the particle spectrum, requiring rather specific BSM
physics. But maybe the price paid is too high? Indeed Pauli finishes
his own comments with the observation:
\begin{quotation}
These requirements are so extensive that it is rather improbable that
they are satisfied in reality.
\end{quotation}
(And Pauli was only concerned with Step 1; the particle masses and
their influence on the cosmological constant. His technique was
completely unable to even begin addressing Newton's constant and
$K_{2/4}$ as influenced by the $k_i$'s and their constraints.)

\section{Frolov--Fursaev: one-loop calculability}
\noindent
Suppose we boldly assume \emph{both} Pauli compensation and Sakharov
one-loop dominance. That is, assume all finiteness constraints:
\be
\str(I) = \str(m^2) = \str(m^4) = 0.
\ee
\be
\str(k_1) = \str(k_1 \; m^2) = 0.
\ee
\be
\str(k_2') = \str(k_4') = 0.
\ee
And in addition, assume all zero-loop (tree level) coefficients are
zero. This is (essentially) the Frolov--Fursaev variant of Sakharov's
proposal.\cite{Frolov} (Note that in the specific model they
implement, Frolov and Fursaev do not impose finiteness/calculability
on $K_2$ and $K_4$, and only impose the first two lines of
constraints.)  Then $\Lambda$, $G$, and $K_{2/4}$ are one-loop
\emph{calculable}:
\be
\Lambda = 
+ {1\over64\pi^2}\;
\str\left[m^4 \ln\left({m^2\over\mu^2}\right)\right] 
+ \hbox{two loops.}
\ee
\be
{1\over G} = 
-{1\over2\pi}\;
\str\left[k_1 \;m^2 \;\ln\left({m^2\over\mu^2}\right) \right]
+ \hbox{two loops.} 
\ee
\be
K_{2/4} = {1\over32\pi^2}\;
\str\left[ k_{2/4}' \; \ln\left({m^2\over\mu^2}\right) \right]
+ \hbox{two loops.} 
\ee
In contrast to Sakharov's original idea,\cite{Sakharov} the cutoff has
now disappeared. Given a particle spectrum that satisfies the
finiteness constraints, the one-loop induced gravitational couplings
are unambiguous. This is a very interesting proposal, but it requires
{\emph{very}} tight interlocking constraints (extended Pauli
compensation) on the particle spectrum to permit it to work.

\section{Renormalizability: invariance constraints}

Suppose we want to be a little less ambitious, perhaps the
interlocking constraints of Pauli compensation seem excessive, perhaps
Sakharov's version of one-loop dominance with an explicit cutoff seems
inelegant: Is there anything interesting you can say just using
``standard'' renormalization theory? 

By this I mean that we are asking that particle physics be at least
one-loop renormalizable in an arbitrary classical background
gravitational field --- note that I am not placing any constraints on
renormalizability of gravity itself. (Indeed at this stage ``gravity''
is a purely descriptive statement about the existence of curved
spacetime and has no dynamics.)  Instead I am imposing the much weaker
condition that particle physics does not violently damage classical
Einstein gravity. The very fact that both particle physics and
classical Einstein gravity work reasonably well on the surface of
planet Earth is good evidence that this is a quite reasonable physical
demand.

Then renormalization in the gravity sector does indeed allow you to
absorb the one-loop divergences into the zero-loop bare quantities ---
but you are allowed to do this once and once only.  Once you have
renormalized, you are not allowed to change your mind about how big an
infinity you dump into the zero-loop bare quantities.

In particular: This means that QFTs can naturally be assigned to
equivalence classes under renormalization of the gravity sector. As
one tunes the QFT coupling constants, so that the QFT goes through a
phase transition or SSB, then the masses in the particle spectrum
change, (and the very content of the particle spectrum changes due to
Higgs particles being eaten by gauge bosons). Then
gravitationally-equivalent QFTs are defined by demanding that the
coefficients of ${\kappa^4}$, ${\kappa^2}$, and
${\log(\kappa^2/\mu^2)}$ in the regulated effective action are not
allowed to change.  (Otherwise you are doing the equivalent of
changing the zero of energy in the middle of the calculation, or
worse.)  That is, as the QFT goes through a phase transition or SSB,
some specific quantities should be held {\emph{invariant}}, and in
particular:
\be
\delta \str(I) = \delta \str(m^2) = \delta\str(m^4) = 0.
\ee
\be
\delta \str(k_1) = \delta \str(k_1 \; m^2) = 0.
\ee
\be
\delta \str(k_2') = \delta \str(k_4') = 0.
\ee
Note that as long as one is working strictly in flat space, the
invariance constraints involving the $\str(k_i)$ would never
appear. The invariance constraints involving $\str(m^n)$ appear only
if you take the net zero of energy seriously --- and it has been
traditional in flat-space QFT to simply ignore the overall net zero
and only ask questions about excitations.

Some of these constraints are automatically satisfied --- (1) Changing
the coupling constants, even if one induces SSB or a phase transition
does not change the net number of fermions or bosons so
$\delta\str[I\,]=0$ automatically. (2) As one goes through SSB a Higgs
scalar is ``eaten'' by a gauge boson which becomes massive. However
$k_1$ for a massive vector is calculated by adding $k_1$ for a
minimally coupled scalar to $k_1$ for a massless vector. Other
particles may acquire mass but do not change their intrinsic
character. Thus $\delta\str[k_1]=0$ provided all the Higgs particles
are minimally coupled scalars [$\,\xi=0\,$]. (3) The same argument
applies to $\delta\str[k_2']$ and $\delta\str[k_4']$, without now any
constraint on the curvature coupling.  Thus the only nontrivial
invariance constraints are
\be
\delta \str(m^2) = \delta\str(m^4) = \delta \str(k_1 \; m^2) = 0.
\ee
We can view these invariance constraints as demanding that for
gravitationally-equivalent QFTs there are strong correlations between
mass changes in the fermi and bose sectors. It is far from trivial to
satisfy these constraints, but it is certainly much easier to do this
than to satisfy the much stronger (extended) Pauli finiteness
constraints.  Provided these invariance constraints are satisfied then
the gravitational couplings have finite
\emph{changes} as one varies the parameters that are one-loop
calculable:
\be
\delta \Lambda = 
+ {1\over64\pi^2}\;
\delta \str\left[m^4 \ln\left({m^2\over\mu^2}\right) \right]
+ \hbox{two loops.} 
\ee
\be
\delta\left({1\over G}\right) = 
-{1\over2\pi}\;
\delta\str\left[k_1 \;m^2 \;\ln\left({m^2\over\mu^2}\right) \right]
+ \hbox{two loops.} 
\ee
\be
\delta K_{2/4} = {1\over32\pi^2} \;
\delta \str\left[ k_{2/4}' \; \ln\left({m^2\over\mu^2}\right) \right]
+ \hbox{two loops.} 
\ee
In this case, if you know the \emph{change} in the particle spectrum,
and can verify that it satisfies the invariance constraints, then you
can calculate (at one-loop) finite changes in $\Lambda$, $G$, and
$K$. (Note these changes are $\mu$-independent.)

To reiterate: these invariance constraints are \emph{not} satisfied by
the standard model (SM) as you modify the Higgs couplings or the
Yukawa couplings.  That is: In the SM coupled to classical gravity
(the curved-space SM), even after you renormalize, all of the
gravitational couplings, ${\Lambda}$, ${G}$, and $K$, are ``unstable''
in the sense that they still suffer infinite shifts whenever the SM
coupling constants vary infinitesimally.  This is one particularly
acute version of the ``cosmological constant problem'' (itself a
variant of the ``hierarchy problem'') demonstrating why ``beyond the
standard model'' (BSM) physics is essential.  It is not just $\Lambda$
that is at risk, the Newton constant $G$ is also subject to
potentially disastrous sensitivity to the coupling constants.  Even if
BSM physics patches up the invariance constraints, one should still
expect finite shifts in $\Lambda$, $G$, and $K$.  Can these be made
compatible with experiment/observation?

Order unity changes in $K_{2/4}$ are not a problem since the
experimental bounds on these constants are so poor. Assuming our
invariance constraints, uncertainties in the Newton constant due to
uncertainties in the coupling constants are of fractional order
$\Delta g \;(M_{EW}/M_{Planck})^2 \approx \Delta g\; 10^{-34}$, and
unlikely to cause immediate concern. It is changes in the cosmological
constant that still prove challenging. It is still distressingly easy
to generate a ``large'' shift in the cosmological constant. (That is,
``large'' as far as cosmology/astrophysics is concerned, consisting of
numerous partially-cancelling terms each of which is individually
$O(M_{EW}^4)$; this is ``small'' from a particle physics perspective
but would still seem to require some fine tuning.)

Suppose now that instead of comparing two different QFTs with
different couplings we want to consider a single definite QFT as a
function of renormalization scale.  Then the the discussion is better
rephrased in terms of $\beta$-functions and anomalous dimensions
($\gamma$-functions).  Define, in the usual particle physics fashion
\be
\beta_g \equiv \mu {\partial g\over\partial\mu}; 
\qquad\qquad 
\gamma_m \equiv {\mu\over m} {\partial m\over \partial \mu}.
\ee
Then one can define differential versions of the invariance constraints:
\be
\str\left[\gamma_m\; m^2 \right] = \str\left[\gamma_m\;m^4 \right] 
= \str\left[k_1 \; \gamma_m\;\;m^2\right] =0,
\ee
though the physics interpretation is now rather different.  These
differential versions of the invariance constraints greatly simplify
the gravitational $\beta$-functions so that
\bea
\beta_\Lambda &=&
- {1\over32\pi^2}\; 
\left\{\str\left[m^4\right] 
- 2 \str\left[\gamma_m \; m^4 \ln\left({m^2\over\mu^2}\right) \right] \right\}
+ \hbox{two loops.} 
\\
\beta_{1/G} &=& 
+{1\over\pi}\;
\left\{\str\left[k_1 \;m^2 \right] 
- \str\left[k_1 \; \gamma_m \; m^2 \ln\left({m^2\over\mu^2}\right) \right] \right\}
+ \hbox{two loops.} 
\\
\beta_{K_{2/4}} &=&  
+{1\over16^2\pi}\;
\left\{ \str\left[ k_{2/4}' \right]
- \str\left[ k_{2/4}' \; \gamma_m \right]  
\right\}   
+ \hbox{two loops.} 
\eea
(Because of the constraints, the apparent explicit $\mu$ dependence is
again an illusion.)  If we now move into a region where the particle
masses are approximately constant $\gamma_m\approx 0$, explicit
integration of the beta functions is trivial: the gravitational
couplings exhibit standard logarithmic running. Indeed the one-loop
regulated formulae for $\Lambda$, $G$, and $K_{2/4}$ now lead to:
\be
\Lambda(\mu) = \Lambda(\mu_0) 
- {1\over64\pi^2}\; \str\left[m^4\right] \ln\left({\mu^2\over\mu^2_0}\right)
+ \hbox{two loops.} 
\ee
\be
{1\over G(\mu)} = {1\over G(\mu_0)} 
+{1\over2\pi}\;\str\left[k_1 \;m^2 \right] \;\ln\left({\mu^2\over\mu^2_0}\right)
+ \hbox{two loops.} 
\ee
\be
K_{2/4}(\mu) = K_{2/4}(\mu_0) 
+{1\over32\pi^2} \str\left[ k_{2/4}' \right]\; \ln\left({\mu^2\over\mu^2_0}\right)
+ \hbox{two loops.} 
\ee
These formulae should be understood with all the usual particle
physics caveats regarding the physical interpretation of the
renormalization point $\mu$. They are useful shorthand for estimating
the strength of the dressed gravitational vertex (classical external
gravity dressed with one-loop matter) in scattering experiments at an
energy scale $\mu$. One should {\emph{not}} blindly insert these
running couplings into the classical field equations.

All in all, it is clear that even quite ordinary straightforward
applications of usual ideas of renormalizability have quite powerful
implications in curved spacetime --- not only can we extract
information regarding classical gravity, but we can turn around and
use classical gravity to suggest interesting constraints on the
particle spectrum.

\clearpage
\section{Variants}
\noindent
You can also construct variations on these four main themes by mixing
and matching. For instance one attractive variant of the original
Sakharov approach is to improve its treatment of the cosmological
constant by using the Pauli finiteness constraints to deal with
$\Lambda$, while reserving Sakharov's cutoff approach for the Newton
constant $G$ and $K_{2/4}$.

Similarly a variant on one-loop calculability (which is closer in
detail, though not in spirit, to what Frolov and Fursaev actually do)
is to use combined Pauli finiteness and Sakharov one-loop dominance on
$\Lambda$ and $G$, but take care of $K_{2/4}$ using standard
renormalization theory.

There are also variants of induced gravity in $1+1$ and $2+1$
dimensions,\cite{Frolov2d,Frolov3d} where the curvature-squared terms
quietly drop out. In contrast in $5+1$ dimensions (or higher) one
would need to keep track of curvature-cubed terms (or higher) as well.
Other possibilities abound, such as the inclusion of external dilaton
fields or string-inspired moduli fields.

\section{Summary and Discussion}
\noindent
One point that all approaches agree on is that one-loop matter effects
lead to shifts in ${\Lambda}$, ${G}$, and $K$.  Depending on your
choices you can interpret these shifts using either ---
\begin{itemize}
\item
{Sakharov}: one-loop dominance.
\item
{Pauli}: one-loop finiteness.
\item
{Frolov--Fursaev}: one-loop calculable.
\item
Renormalization: calculable one-loop \emph{changes}.
\end{itemize}
Whatever your viewpoint in this regard, the key point is this:
\emph{Even if Einstein gravity is not there at zero loops, it will
automatically be generated at one-loop.}

Thus we have seen how Sakharov's ideas have been very fruitful --- and
how they naturally lead in a number of different directions.  There
are still many interesting questions to answer just at the level of
QFT and particle phenomenology. (Without requiring any particularly
radical reassessment of our world-view.) In particular the finiteness
constraints (or even the much weaker invariance constraints) place
very interesting restrictions on the allowable spectrum of particles
in BSM physics.

But Sakharov also teaches us a lot about gravity: Suppose you have a
manifold that is ``free to flap in the breeze'', on which you proceed
to construct a QFT. Then the geometry of that manifold serves as a
classical external field, and the effective action is a function of
that geometry. Extremizing the effective action in the usual manner,
because it automatically contains the classical gravitational action,
automatically leads to semiclassical quantum gravity, at least at the
one-loop level. (By semiclassical gravity I mean that the classical
Einstein equations, plus curvature-squared corrections, are coupled to
the expectation value of the quantum stress-energy tensor for all the
quantized matter fields.)

The important point here is that gravity was never put into the
quantum theory, and gravity was never quantized in any way shape or
form. Nevertheless, classical gravity (meaning the inverse square law
implicit in the Einstein equations and Einstein--Hilbert action)
automatically emerges in the semiclassical limit. If you couple this
with the observation that the only actual experiments we can
(currently) perform with gravity also inhabit this same semiclassical
realm, is there any real need to quantize gravity itself?

Most physicists would still answer this question in the affirmative
--- but then Sakharov's ideas still have considerable impact: If we
consider any candidate theory for quantum gravity (brane models,
quantum geometry, lattice quantum gravity) then Sakharov's scenario
tells us that trying to derive the inverse square law from first
principles is not the difficult step. Deriving the inverse square law,
and all of Einstein gravity is in fact automatic once you have
demonstrated the existence of Lorentzian manifolds --- reasonably
large reasonably flat arenas on which to set up a low-energy QFT.

\nonumsection{Acknowledgments}
I wish to thank Larry Yaffe for helpful comments and criticisms. This
research was supported by the US DOE.


\end{document}